# Bridging Machine Learning and Clinical Diagnosis: An Explainable Biomarker for ß-Amyloid PET Imaging

János Barbero[1], Ana Franceschi[2], Luca Giliberto[1], Patrick Phuoc Do[3], David Petrover[2], Jack Nhat Truong[4], Sean Clouston[5], Nha Nguyen[6], Marc Gordon[2], An Vo[1]

[1]The Feinstein Institutes for Medical Research, Manhasset, NY, [2]Donald and Barbara Zucker School of Medicine at Hofstra/Northwell, Hempstead, NY, [3]University of Massachusetts Amherst, Amherst, NY, [4]Adelphi University, Garden City, NY, [5]Renaissance School of Medicine at Stony Brook University, Stony Brook, NY, [6]Albert Einstein College of Medicine, New York, NY

## Introduction:

[18F]-florbetaben positron emission tomography (PET) imaging is an established marker of ß-Amyloid (Aß) that is being increasingly used to assess Aß deposition in Alzheimer's Disease (AD) [Sabri et al., 2015]. However, evaluation of these scans currently relies on expert interpretation. While prior machine learning classifiers for [18F]-florbetaben imaging exist [Lee et al., 2021], they lack the transparency needed for clinical decision- making. Our work bridges this gap, introducing an explainable machine learning biomarker for assessing Aß deposition in a more interpretable and clinically relevant manner.

## Methods:

We analyzed 163 [18F]-florbetaben PET scans acquired at our institution as part of a retrospective analysis [Franceschi et al., 2023]. Each scan was interpreted by two independent expert readers to assess Aß+ status, blinded to each other's assessments. We segmented and registered the scans into Montreal Neurological Institute (MNI) space using Statistical Parametric Mapping (SPM) 12, parcellated them using the Automated Anatomical Labeling (AAL) atlas [Tzourio-Mazoyer et al., 2002], then computed mean expression scores in each region normalized by mean cerebellar expression using a custom program in Matlab R2023a. We used the regional loadings to train a Cubic Support Vector Machine (SVM) classifier and tested the model using 5-fold cross-validation. To elucidate the model's decision-making process in aggregate, we employed local interpretable model-agnostic explanations (LIME) [Ribeiro et al., 2016], projecting the most influential features back onto the AAL atlas for visualization.

## Results:

Our cohort comprised 163 patients with [18F]-florbetaben imaging obtained from 2016 to 2018 (84M, 83F; 76.1 ± 6.8yo). There was high agreement (92.0%, k=0.84) between the readers, with 90 scans interpreted as positive and 60 as negative by both. The Cubic SVM classifier demonstrated robust performance, achieving 92.0% accuracy and 90.5% precision under 5-fold cross-validation. The model was highly sensitive (95.6%) and specific (87.7%) to Aß+ status, with an area under the curve (AUC) of 95.0% and an F1 score of 93.0% (Fig. 1). LIME analysis revealed that the model identified critical regions impacting Aß+ status, with notable contributions from the inferior frontal cortex, cuneus, olfactory cortex, postcentral gyrus, supramarginal gyrus, temporal pole, thalamus, and pallidum (Fig. 2). In our findings, amyloid deposition within the temporal pole emerged as the most significant factor in confirming Aß positivity. Conversely, the absence of amyloid deposition in the cuneus was identified as the key indicator for ruling out Aß+ status.

## Conclusions:

This study presents a novel, explainable machine learning-based biomarker for assessing Aß+ positivity based on [18F]-florbetaben PET scans. The model showcases high accuracy, paralleling expert interpretation levels, and provides critical insights into the brain regions most affected by amyloid deposition. Unlike other black-box classifiers, our model transparently delineates the specific brain regions influencing its classifications. This clarity is crucial in a clinical setting, where the rationale behind diagnostic decisions must be both understandable and justifiable. In the context of emerging amyloid-focused disease-modifying therapies, the transparency and precision of our model make it a robust tool for patient selection, assessment of treatment efficacy, and monitoring of disease progression in a clinically interpretable manner. Finally, by elucidating the relationship between overall Aß+ status and specific brain regions, our approach holds promise for novel insights into the correlation between the spatial distribution of Aß and the clinical manifestations of Alzheimer's Disease.

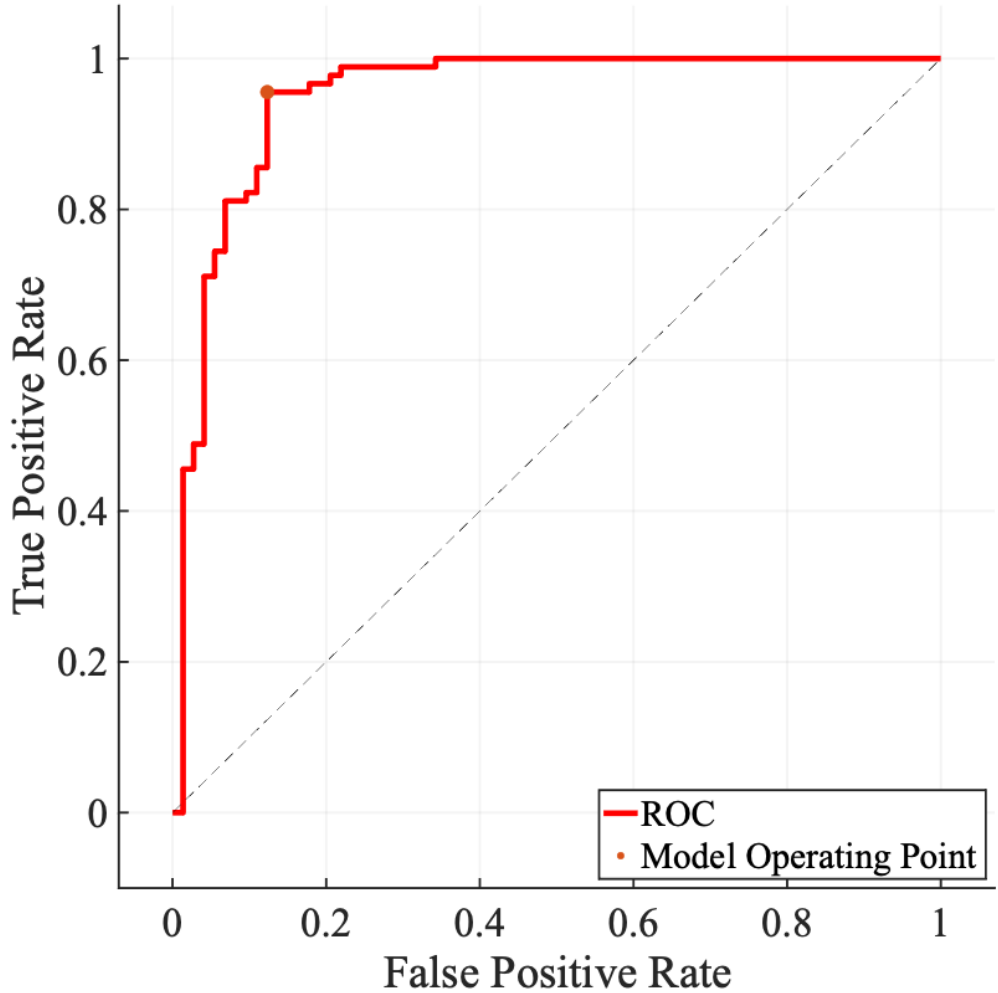

**Figure 1**: Receiver operating characteristic curve of a Cubic Support Vector Machine classifier to evaluate Aβ+ in [$^{18}$F]-florbetaben PET scans. The model is highly sensitive (95.6%) and specific (87.7%), with a total AUC of 95.0%.

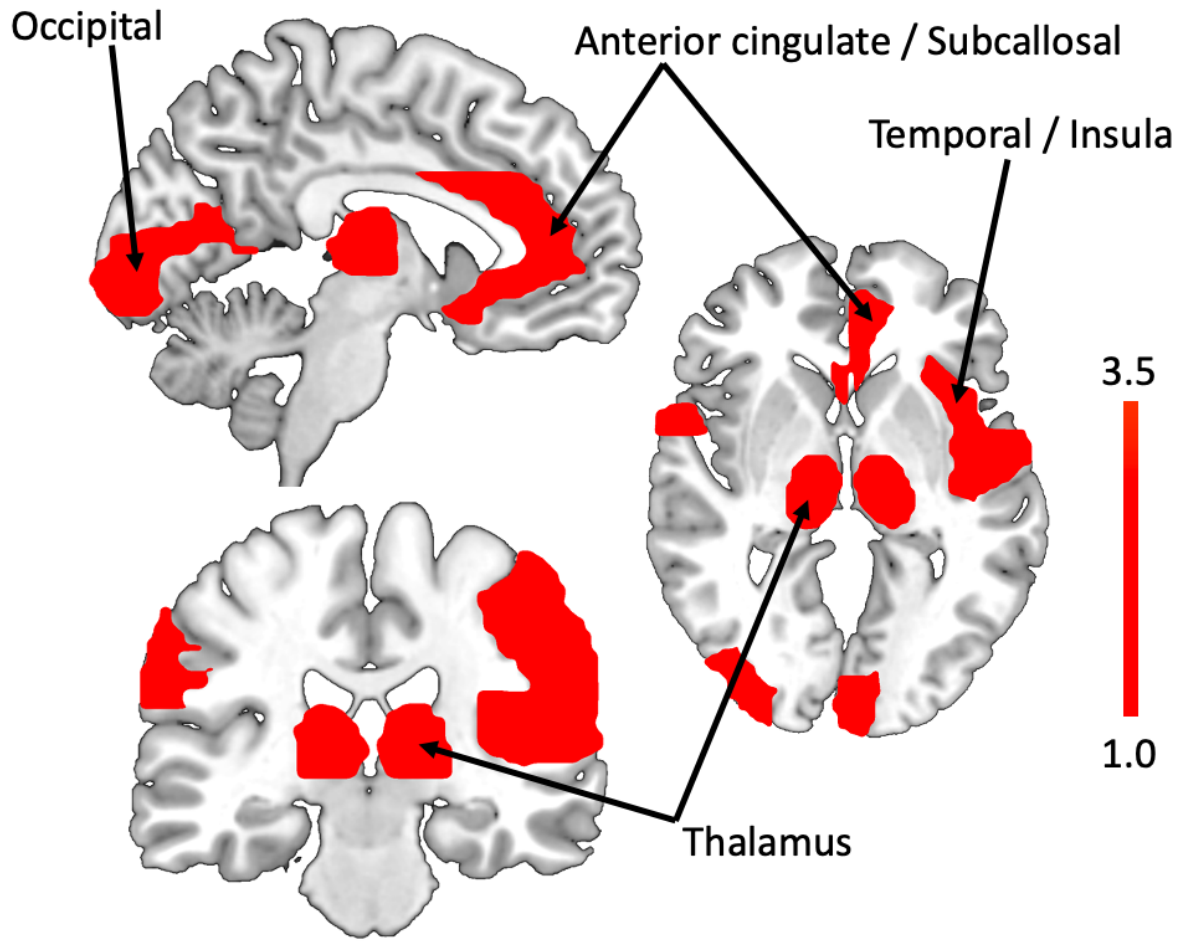

**Figure 2**: Local interpretable model-agnostic explanations (LIME) region weights for β-Amyloid (Aβ) positive predictions. Amyloid burden in regions that contributed significantly to predicting Aβ+ status include the temporal cortex, inferior frontal and orbitofrontal cortices, insula, postcentral gyrus, supramarginal gyrus, and thalamus.